\documentclass[lineno]{biometrika}

\usepackage{amsmath}
\usepackage{epstopdf}

\usepackage{caption}

\usepackage{times}
\usepackage{bm}
\usepackage{natbib}
\usepackage{bbm}
\usepackage[plain,noend]{algorithm2e}

\usepackage{graphicx}
\usepackage{subfigure,color}

\usepackage[normalem]{ulem}

\makeatletter
\renewcommand{\algocf@captiontext}[2]{#1\algocf@typo. \AlCapFnt{}#2} % text of caption
\def\@algocf@capt@plain{top}
\renewcommand{\algocf@makecaption}[2]{%
  \addtolength{\hsize}{\algomargin}%
  \sbox\@tempboxa{\algocf@captiontext{#1}{#2}}%
  \ifdim\wd\@tempboxa >\hsize%     % if caption is longer than a line
    \hskip .5\algomargin%
  \parbox[t]{\hsize}{\algocf@captiontext{#1}{#2}}% then caption is not centered
 \else%
   \global\@minipagefalse%
   \hbox to\hsize{\box\@tempboxa}% else caption is centered
 \fi%
  \addtolength{\hsize}{-\algomargin}%
}
\makeatother

\def\halftext{.5\textwidth}

\newcommand{\iidsim}{\stackrel{\mbox{\tiny{iid}}}{\sim}}
\newcommand{\defeq}{\stackrel{\mbox{\tiny{def}}}{=}}
\begin{document}

\jname{Biometrika}
%% The year, volume, and number are determined on publication
\jyear{2013}
\jvol{xx}
\jnum{x}
%% The \doi{...} and \accessdate commands are used by the production team
%\doi{10.1093/biomet/asm023}
%\accessdate{Advance Access publication on 27 December 2013}
\copyrightinfo{\Copyright\ 2013 Biometrika Trust\goodbreak {\em Printed in Great Britain}}

%% These dates are usually set by the production team
%\received{December 2013}
%\revised{December 2013}

%% The left and right page headers are defined here:
%\markboth{A. C. Davison, R. Gessner \and D. M. Titterington}{Biometrika style}

\markboth{D. N. VanDerwerken \and S. C. Schmidler}{Biometrika style}

%% Here are the title, author names and addresses
\title{Parallel Markov Chain Monte Carlo}

\author{D. N. VanDerwerken, S. C. Schmidler}
\affil{Department of Statistical Science, Duke University, Durham, North Carolina 27708, U.S.A. 
\email{dnv2@stat.duke.edu} \email{schmidler@stat.duke.edu}}
\maketitle

\begin{abstract} 
Markov chain Monte Carlo is an inherently serial algorithm.  Although likelihood calculations for individual steps can sometimes be parallelized, the serial evolution of the process is widely viewed as incompatible with parallelization, offering no speedup for samplers which require large numbers of iterations to converge to equilibrium.  We provide a methodology for parallelizing Markov chain Monte Carlo across large numbers of independent, asynchronous processors.  Our approach uses a partitioning and weight estimation scheme to combine independent simulations run on separate processors into rigorous Monte Carlo estimates. The method is originally motivated by sampling multimodal target distributions, where we see an exponential speedup in running time.  However we show that the approach is general-purpose and applicable to all Markov chain Monte Carlo simulations, and demonstrate speedups proportional to the number of available processors on slowly mixing chains with unimodal target distributions. The approach is simple and easy to implement, and suggests additional directions for further research.
\end{abstract}


\section{Introduction}
Markov chain Monte Carlo methods are the dominant computational tool in Bayesian statistics, and their emergence is commonly credited with the meteoric rise of modern Bayesian analysis.  Markov chain Monte Carlo renders computationally tractable many important applied Bayesian analyses that were analytically intractable, and therefore impractical, 25 years ago.  However, the ensuing spread of Bayesian statistical modeling has led to new computational challenges for Bayesians, as models become more complex and higher-dimensional, and both parameter sets and data sets become orders of magnitude larger.

Additionally, the computing world has changed: the steady march of Moore's law - the exponential growth in processor speeds that has continually expanded the applicability of Markov chain methods - is gone.  Computer manufacturers have shifted from a model of increasingly faster CPUs to one of growing parallelism: multi-core platforms, cluster computing, and the massive parallelism of GPUs.  Unfortunately, Markov chain Monte Carlo is an inherently serial process: reliance on iterative convergence to equilibrium and long-run averages means that the $n$ steps of a Markov chain cannot be computed in parallel because step $i$ depends on steps $1,\ldots,i-1$.  Although some effort has been put into parallelizing individual Markov chain steps, the simulation of chains requiring large numbers of iterations to converge is generally viewed as unparallelizable.  Running multiple chains in parallel does not lead to significant speedups when convergence is slow, since each must be run long enough to equilibrate \citep{Geyer:1992}.  Similar arguments apply to other dynamical simulation problems where interest is in the equilibrium behavior of systems with relaxation times, such as molecular dynamics simulation.

Here we challenge this view of Markov chain Monte Carlo as inherently serial, by introducing to our knowledge the first general-purpose method for parallelizing an arbitrary Markov chain Monte Carlo algorithm.  Our approach provides a simple yet rigorous methodology for combining the output of an arbitrary number of completely independent - and therefore parallelizable - copies of the Markov chain, in a way that achieves a speedup in estimating equilibrium properties that is proportional to the number of available processors. This provides a simple conceptual basis for asynchronous parallelization of general purpose Markov chain Monte Carlo algorithms, that can be applied directly to chains generated by existing software packages such as BUGS, or complex multi-chain samplers such as parallel tempering or adaptive Markov chains.  Our results indicate a speedup which is roughly linear in the number of processors in unimodal target distributions; while multimodal target speedups can be significantly larger, growing exponentially in the problem dimension. Many directions for improvements in efficiency remain to be explored, and we feel this is an exciting area for further research.

\section{Previous work on parallelization} 
Previous attempts to parallelize Markov chain Monte Carlo have been much more limited, most commonly to parallelize expensive likelihood evaluations, e.g. \citet{Feng:2003}. This can provide dramatic speedups in some problems, but is problem-specific.  It also requires careful processor synchronization, achievable only on dedicated clusters with high-speed connectivity, without which slow-downs over serial computing are common.  Similar arguments apply to proposing several moves in parallel, e.g. H. M. Austad's 2007 Norwegian University of Science and Technology Masters thesis, or precomputing acceptance ratios in parallel \citep{Brockwell:2006}.  However since they reduce the time-per-step, such methods, when available, can be used in concert with our approach to provide even further improvements.  Closer in spirit to our approach, \citet{Craiu:2009} use an adaptive Metropolis method and adapt chain-specific proposal covariances using the pooled draws from several parallel chains, and \citet{Wang:2012b} use parallel Wang-Landau processes to explore the posterior and adapt a Metropolized independence proposal distribution \citet{Ji:2007}.  Finally, \citet{Lee:2010} show how graphics cards offer massively parallel simulation for certain classes of `population-based' algorithms, including parallel tempering and sequential Monte Carlo methods.

All of these previous methods require the individual chains to reach the stationary distribution in order for inference to be valid, and thus cannot significantly reduce the number of serial steps the algorithm must take.  When convergence to equilibrium is slow, these methods will not help; e.g. in the presence of significant multimodality, the samplers may not reach stationarity in any practical timescale. By contrast, the method we propose here is especially designed for sampling from multimodal distributions, though it applies readily in unimodal settings as well.  \citet{Rosenthal:2000} gives several examples of Monte Carlo methods that are `embarrassingly parallelizable' by concurrent sampling of independent trajectories, when mixing of individual chains is known to be very fast.  Our approach is designed to make any Markov chain Monte Carlo algorithm embarrassingly parallelizable, by requiring only that each chain mix locally, and learning the marginal probability of local regions to allow results to be rigorously combined.

\section{Methods}
\subsection{Overview of our approach}
The basic idea of parallelization is to break a problem into pieces that can be solved independently - preferably asynchronously - and recombined into a full solution.  For integrating functions with respect to a probability measure $\pi(\theta)$ on a metric space $\Theta$, one possibility is to partition the space $\bigcup_{j=1}^J \Theta_j$, integrate within each partition element, and sum the results.  For numerical quadrature this is easily done in low dimensions; however, for fixed error the number of partition elements required grows exponentially in the dimension of $\Theta$.  This is why Markov chain Monte Carlo methods, which spend their budgeted function evaluations in the relevant parts of the space, are preferred for most Bayesian computations.  However, as noted above, Markov chain methods are inherently serial.  Researchers are therefore often faced with a choice between two classes of algorithms, both requiring an exponential number of function evaluations.  Here we outline a method that attempts to combine the best of both worlds, spending its time only in regions with significant probability while enabling parallel evaluation in distinct regions.  We achieve this using an adaptive partitioning scheme based on Markov chain sample trajectories.

We first describe the general scheme for a given partition.  Suppose for each $j$ we run a Markov chain Monte Carlo algorithm $\theta_1^{(j)},\ldots,\theta_{n_j}^{(j)}$ on the target distribution restricted to $\Theta_j$, that is $\pi_j(\theta) \defeq \pi(\theta)\mathbf{1}_{\Theta_j}(\theta)/w_j$ where $w_j = \int_{\Theta_j} \pi(\theta)\nu(d\theta)$.  For $\pi$-measurable functions $f$, the ergodic averages $\hat{u}_{j,n}= n_j^{-1}\sum_{i=1}^{n_j}f(\theta_j^{(i)})$ provide estimators converging to $E_{\pi_j}(f) = \int_{\Theta_j} f(\theta) \pi_j(\theta)\nu(d\theta)$ respectively for $j\in\{1,\ldots,J\}$ as $n_j\rightarrow \infty$.  These can be combined with estimated weights $\hat{w}_{j,n}$ converging to $w_j$, to obtain
\begin{equation*}
\hat{\mu}_n = \sum_{j=1}^J \hat{w}_{j,n}\hat{\mu}_{j,n}
\end{equation*}
converging to $\mu=E_\pi(f)$.  If the $\hat{u}_{j,n}$'s and $\hat{w}_{j,n}$'s are unbiased and independent, $\hat{\mu}_n$ is unbiased.

This approach has two shortcomings: first, it requires a number of independent simulations, and thus processors, equal to the size of the partition; this may grow exponentially in dim($\Theta$). Second, the rejection often needed for the restriction doesn't permit easy evaluation of transition kernel densities, required below. In addition, estimating the relative weights $w_j$ with which they should be combined requires care.

The first shortcoming is addressed by noticing that in the above scenario each Markov chain is simply an estimator of $E_{\pi_j} f$, and we need not have a one-to-one correspondence between chains and estimators.  The required set of estimators can be produced by a much smaller, i.e. non-exponential, number $L$ of independent Markov chains by $\hat{\mu}_{j,n} = n_j^{-1} \sum_{l=1}^L \sum_{k=1}^{K_l} f(\theta_{lk})\mathbf{1}_{\Theta_j}(\theta_{lk})$ where now $n_j = \sum_{l=1}^L \sum_{k=1}^{K_l} \mathbf{1}_{\Theta_j}(\theta_{lk})$ is the number of draws in $\Theta_j$ across all chains.  This also solves the second problem, since the chains need no longer be restricted and can cross between partition elements.

\subsection{Estimating the weights}
\label{Sec:Weights}
Let $\pi(\theta) = g(\theta)/c$ where $g$ is the unnormalized target density, and let $c_j = \int_{\Theta_j}g(\theta)\nu(d\theta)$.  Estimating $c_j$ is equivalent to estimating the normalizing constant of the density $g_j(\theta) = g(\theta) \mathbf{1}_{\Theta_j}(\theta)$; a number of techniques exist for estimating normalizing constants.  Importantly, the restriction to $\Theta_j$ allows us to rule out pathologies that can make such estimation unreliable.

{\it Approach 1: Markov chain output} \; We can estimate $c_j$ directly from the trajectory of the Markov chain with target distribution $\pi_j$, using techniques for estimating marginal likelihoods \citep{Gelfand:1994b,Chib:1995,Chib:2001,Raftery:2007}, though care is needed \citep{Wolpert:2012}.
Most of these can be interpreted as bridge sampling estimators \citep{Meng:2002}, suggesting further improvements.  

{\it Approach 2: Adaptive importance sampling} \; Often a more efficient estimator can be obtained by using the Markov chain draws to construct an approximation $q_j$ to the restricted target distribution $\pi_j$, and then drawing $\theta_t \iidsim q_j$ to form an unbiased importance sampling estimate $\hat{c}_j = T^{-1}\sum_{t=1}^T g(\theta_t)\mathbbm{1}_{\Theta_j}(\theta_t)/q_j(\theta_t)$.  As usual $q_j$ must satisfy $\mbox{supp}(q_j) \supset \mbox{supp}(\pi_j)$ and $\lambda_j^* = \sup{\pi_j/q_j} < \infty$.  Since $q_j$ need only approximate the target locally, on the partition element, $\lambda_j^*$ is typically much smaller than if applied to the global target distribution. As above, many variations exist: importance sampling estimates of $w$ or $1/w$, bridge sampling \citep{Meng:1996}, path sampling \citep{Gelman:1998}, and variance reduction techniques can help. More generally we may use a sequence $q_{j,t}$ of distributions, such as a Markov chain $\theta_t \mid \theta_{t-1} \sim q_j(\theta_{t}\mid\theta_{t-1})$. If the Markov chain has limiting distribution $\pi_j$ this is similar to Approach 1 without requiring restriction, but it need not have. A single approximating normal or $t$-distribution with moments matched to the samples may perform poorly when $\pi_j$ is skewed or multimodal. \citet{Ji:2007} show how a finite mixture of normal or $t$-distributions can be estimated in real time, allowing for concurrent sampling, although a single component often suffices if the partition is fine enough. We use `sample' and `trajectory' to denote a set of independent and conditional draws, respectively.  Averaging across $n$ independent $\hat{c}_j$'s will decrease the variance at a rate of $n^{-1}$.

\subsection{Combining the weights} 
{\it Ratio estimation:} The identity $ c_j = w_jc$ naturally suggests the estimator $\hat w_{j,n} = \sum_{i=1}^n \hat{c}_j^{(i)}/\big[\sum_{i=1}^n \sum_{k=1}^J \hat{c}_k^{(i)} \big]$, which is consistent, but not unbiased, for $w$. Other ratio estimators \citep{Tin:1965} may improve efficiency, permitting a reduction in $n$.  

{\it Pseudo-marginal sampler:}\; We have also explored using the $\hat{c}_j$'s to simulate a finite state Markov chain with stationary distribution $(w_1,\ldots,w_J)$ obtained by the pseudo-marginal approach of \citet{Andrieu:2009}.
Empirically, we've found that for a given $T$ and number of pseudo-marginal iterations, the mean square error of the $\hat w_j$'s obtained this way outweighs the bias of direct estimation using the same set of $\hat{c}_j$'s as described above. 
 
\section{Illustrative Examples}
\subsection{Multimodality}
Two of the most common causes of slow Markov chain convergence are multimodality, where a random-walk converges quickly to a local mode of the target, but takes a long time to escape to other significant modes; and dimensional dependence, which dramatically slows component-wise algorithms such as Gibbs samplers, and makes efficient joint proposals difficult to find.   We show that parallelization approach can yield significant speedups in problems of each type.

When the posterior is rough or multimodal - often due to nonlinearity in the likelihood, model misspecification, or model selection - Markov chains can get trapped in local modes with long waiting times for transitions between modes. This behavior can often be identified by simulating chains from distinct initializations, and in some cases it may even be possible to locate all relevant modes by optimization techniques. In this case the behavior of independent chains running in parallel on different processors is easily pictured, with each chain exploring the basin of the local mode to which it is attracted.  Such cases seem ideal for parallelization, but with no clear way to combine the output of such independent simulations to obtain consistent estimators, current algorithms require waiting for each simulation to converge to equilibrium across all modes, which can take exponentially long \citep{Woodard:2007b}.  The approach presented here represents a solution to this problem which can completely remove the waiting times for crossing between modes, leaving only the relatively short within-mode equilibration times.

{\it Mixture of normals:} As a simple model for multimodal target distributions, consider sampling from a mixture of four bivariate normal distributions with weights $w = (0.02, 0.20, 0.20, 0.58)$; means $\mu_1 = (3,3)$, $\mu_2 = (7,-3)$, $\mu_3 = (2,7)$, and $\mu_4 = (-5,0)$; and corresponding covariance matrices $\Sigma_1 = [1,.2; .2,1]$, $\Sigma_2 = [2,-.5;-.5,.5]$, $\Sigma_3 = [1.3,.3;.3,.4]$, and $\Sigma_4 = [1,1; 1,2.5]$ using a Langevin diffusion $d\theta_t = \sigma^2/2 \nabla \log \pi(\theta_t)dt + \sigma dW_t$ where $W_t$ is standard Brownian motion and $\sigma^2 \ll 1$. Although Metropolis-adjusted Langevin is often preferred, evaluating trajectory densities for Metropolis steps is difficult and requires Approach 2; we defer this to below.

Like other random-walk chains, the Langevin diffusion quickly explores the mixture component nearest its initial state, but mixes poorly between modes.  To demonstrate our parallelization method, we defined a partition using Voronoi cells centered at the four modes.  A diffusion was initialized in each mode and run for 25,000 iterations, requiring $\sim90$s on four 3.40 GHz processors. In the same time, four other processors generated four sets of $n\approx5000$ trajectories of length $T=5$ and corresponding $\hat{c}_j$'s, one set per mode, initialized independently from a $t_4$-distribution centered at the mode with covariance the inverse Hessian of $\log \pi(\theta)$. These short trajectories used $t_4$ perturbations in place of Brownian motion to ensure finite variance of the importance sampling estimate. The estimated weights were 0.02, 0.23, 0.20, and 0.55.

\begin{figure}
\parbox{\halftext}{
\figurebox{15pc}{15pc}{}[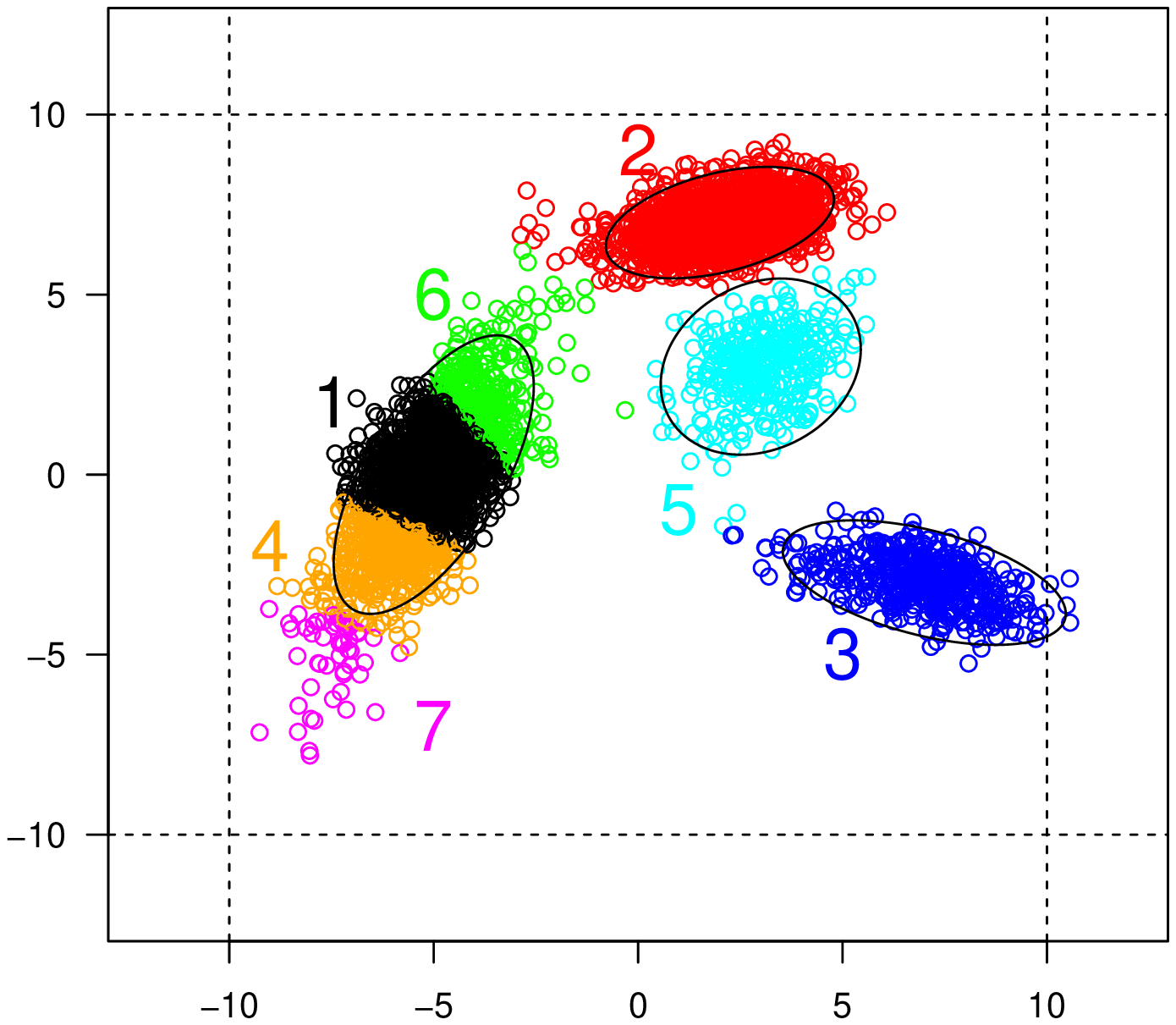]
}
\hfill
\parbox{\halftext}{
\figurebox{15pc}{15pc}{}[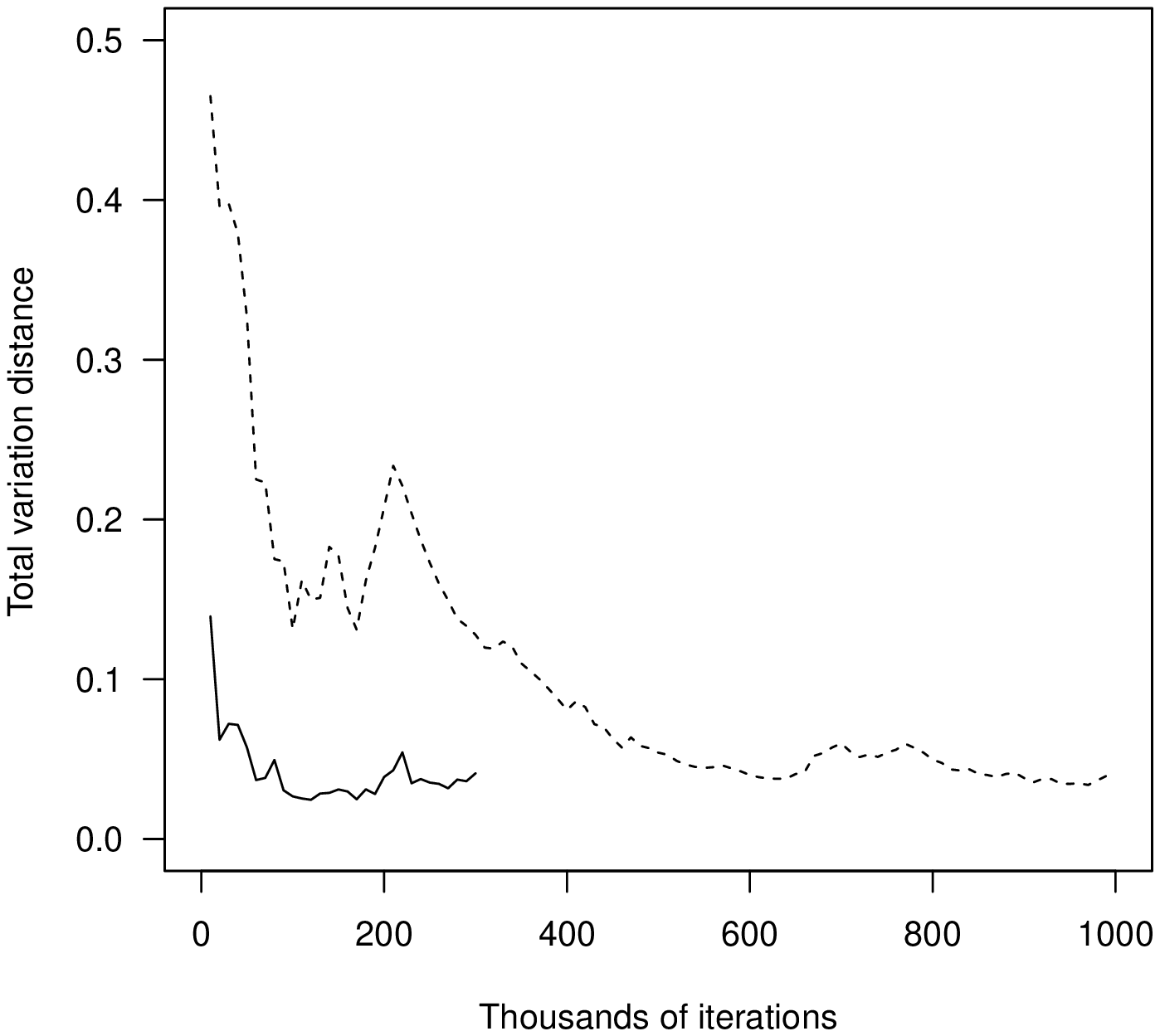]
}
\caption{ (Left) Colors show results of clustering 10 Langevin chains initialized uniformly inside the box (dashed lines);
ellipses show 95\% contours for component densities. (Right) Parallel Gibbs sampling using 10 chains (solid line) speeds up convergence on multivariate probit regression model by approximately $20\times$ as compared with regular Gibbs sampling (dashed line).}
\label{Fig:ClusteringProbit}
\end{figure}

{\it Adaptive partitioning:} The assumption of known modes was for illustrative purposes only, and 
can be avoided using an adaptive partitioning scheme to handle as many modes as are discovered by the multiple chains.  As always we can make no guarantees about discovering all possible modes, but this is not our focus here.  Suppose we believe 10 random initializations sufficient to locate all relevant modes or indicate more are needed. We initialized 10 chains uniformly at random on $(-10,10)\times (-10,10)$ and ran each for 25,000 iterations in parallel on 10 processors.  Following a burn-in of 250 the first 1,000 draws were clustered into modes. We used the technique of \citet{VanDerwerken:2013} with $\alpha=0.01$, $\epsilon^2 = 9$, which groups all points within 
$\epsilon$ of the highest density remaining draw, and repeats until $(1-\alpha)100\%$ ($\alpha>0$) of draws have been clustered, then reallocates draws to the nearest center inducing a Voronoi partition. This resulted in a partition of seven elements, see Fig.~\ref{Fig:ClusteringProbit}.  

Repeating the above steps with this partition gives similar results.  Alternatively we can use Approach 2 of Section~\ref{Sec:Weights}, drawing independently from the $t_4$-distributions to perform importance sampling.  If the $t_4$ reasonably approximates $\pi_j$ this can be significantly more efficient: we obtained 5000 $\hat{c}_j$'s from samples of size $T=5$ in $18$s vs. $90$s for simulating the diffusions. The clustering and importance sampling takes less than half the time required for the parallel 25,000 chains, enabling the weights to be estimated in parallel before the sampling completes.  As seen in Fig.~\ref{Fig:ClusteringProbit}, partition element $\Theta_5$ corresponds to the first mixture component $C_1$; $\Theta_7 \approx C_2$; $\Theta_2\approx C_3$; and $\Theta_1 \cup \Theta_4 \cup \Theta_6 \cup \Theta_7 \approx C_4$. The estimated weights are $\hat{w} =(.378, .201, .201, .105, .020, .093, .002)$; summing gives estimated component weights $(.020, .201, .201, 0.578)$ which are nearly exact. Thus our method does not require exact identification of modes or restriction of local samplers to a single partition element.

{\it Higher dimensions:} A more difficult mixture target in $p=10$ dimensions was constructed by drawing four component-means uniformly on $(-10,10)^p$, random correlation matrices $L^TL$ with $L\sim\mathcal{MN}_{p\times p}(0,I_p,I_p)$, and component weights $\sim$Dirichlet$(1,1,1,1)$. We ran 20 parallel Metropolis chains with independent normal proposals for 100,000 iterations, independently initialized uniformly on $(-10,10)^p$. Scale parameters were tuned automatically during the first $1000$ iterations, which were discarded as burn-in.  After 49,000 post-burn-in iterations, draws were clustered as above ($\epsilon^2=p=10$, $\alpha=0.05$, columns normalized), yielding four partition elements. $t_4(m_j,S_j)$ distributions were constructed, with $m_j$ the cluster centers and $S_j$ the empirical covariance of draws in element $j$, for independent importance sampling with $T=100$, $n=1000$ to obtain $\hat{c}_j$'s. Total variation distance between the estimate and 'true' weights, given by the proportion of 100,000 independent draws falling in each partition element, was 0.0024. The component-averaged absolute error between the parallel estimate of the expectation, constructed from the above weights and the 99,000 available post-burn-in draws, and the true expectation was 0.17. To test sensitivity to the partition, we reduced $\epsilon^2$ to 9 and repeated; the resulting partition has eight elements and the total variation distance between the weight vector and the truth increases to 0.0074 while the average absolute error decreases to 0.12. The increased weight error is attributable to more, smaller weights being estimated; the decreased overall expectation error is likely due to better mixing within the smaller partition elements. This highlights a tradeoff between estimating weights and within-partition-element expectations.

\subsection{Dependence in Gibbs samplers}
Although easily visualized in such cases, our parallelization approach is not limited to multimodal distributions, and can also be used to speed up convergence of Markov chains that mix slowly for other reasons, such as correlation in Gibbs samplers.

{\it Probit regression:} Consider a probit regression model assigning probabilities $1-\Phi(\beta X)$ and $\Phi(\beta X)$ for a response $Y\in \{0,1\}$ with a single covariate $X$. The posterior distribution for $\beta$ after $n$ observations is proportional to $\pi_0(\beta)\prod_{i=1}^n \Phi(\beta  X_i)^{y_i}\{1-\Phi(\beta X_i)\}^{1-y_i}$.  We simulated $N=2000$ covariate and observation pairs, with $X \sim \text{Bern}(\frac{1}{2})$ and $\beta = 5/\sqrt{2}$, and performed inference under a vague prior ($\pi_0(\beta) = N(0,10^2)$).  This model was also studied by \citet{Nobile:1998} and \citet{Imai:2005}, who noted that the traditional Gibbs sampler \citep{Albert:1993} mixes slowly, with autocorrelation exceeding $0.999$.

We applied our parallelization approach to this Gibbs sampler by running 10 parallel chains initialized uniformly on $(0,20)$ for 50,000 iterations each. We formed the Voronoi partition with centers given by the deciles from the 500,000 pooled draws. Weights were estimated via Approach 2 of Section~\ref{Sec:Weights} with the instrumental distribution for element $j$ being $N(m_j,2s_j)$ for $m_j$ and $s_j$ the mean and standard deviation of all draws in partition element $j$, the factor 2 included to ensure slightly heavier tails. We used $n=500$ and $T=10$.   

Again we calculated total variation distance to the ``true'' posterior of 200,000 independent rejection sampling draws, on a fine discretization $(-\infty, 0] \cup \left\{\cup_{i=1}^{50}((i-1)/2, i/2]\right\} \cup (25,\infty)$, giving distance 0.075. The distance for a single serial Gibbs sampler does not reach 0.075 until 1.2 million iterations. Thus the parallelized Gibbs sampler achieves the same accuracy with less than half as many draws, and more than a factor of 20 speedup due to parallelization.

{\it Multiple covariates:}
Similar results are obtained for higher dimensions.  For $N=500$ observations of $p=8$ covariates drawn from (const.(1), Bern($\frac{1}{2}$), U(0,1), N(0,1), Exp(1), N(5,1), Pois(10), N(20,25)) and $\beta = (0.25, 5, 1, -1.5, -0.1, 0,0,0)^T$, we ran 1M iterations of a serial Gibbs sampler initialized at zero and 300k iterations each for 10 parallel samplers initialized independently from U(-10,10) for $\beta_{1-4}$ and U(-0.3,0.3) for $\beta_{5-8}$. Partitioning used $\epsilon^2=p$ for normalized dimensions, instrumental distributions were $t_4(m_j^*,S_j)$ for $m_j^*$ and $S_j$ the empirical mode and covariance in element $j$, and weight estimation used $n=500$, $T=50$. $\beta_2$ is by far the slowest to converge with autocorrelation $>0.999$, all others $<0.95$, so we compare the sampled marginal distribution for $\beta_2$ with the `true' marginal obtained from 5M Metropolis-Hastings samples with autocorrelation only $0.95$. Fig.~\ref{Fig:ClusteringProbit} shows variation distance calculated using discretization $\{(\infty,3.5],(3.5,4.0],\ldots,(14.5,15.0],(16,17],\ldots(24,25],(25,\infty)\}$ vs. iteration. Using convergence threshold  0.10 \citep{VanDerwerken:2013}, the parallelized Gibbs again sampler converges about 20 times faster. 

\section{Loss of Heterozygosity Data Example}
We applied our technique to parallelize the Metropolis-Hastings chain for the loss of heterozygosity model from the Seattle Barrett Esophagus research project \citep{Desai,Craiu:2009,Bai:2011}.  Loss of heterozygosity is a genetic change undergone by cancer cells during the course of disease, and chromosomal regions with high rates of loss are hypothesized to contain regulatory genes. The data contain 40 observations of number of instances of loss ($X$) and sample size ($N$) pairs. Loss frequencies are modeled using the mixture model:
\begin{equation*}
X_i \sim \eta\,\text{Binomial}(N_i, \pi_1) + (1-\eta)\text{Beta-Binomial}\,(N_i, \pi_2, \gamma),
\end{equation*}
the probabilities of LOH in the binomial and beta-binomial groups, respectively, and 
where $\gamma$ controls overdispersion in the beta-binomial. Independent U($-30$,$30$) priors are assigned to $\gamma$ and the logits of $\eta$, $\pi_1$, and $\pi_2$.  The likelihood function is 
\begin{equation*}
\prod_{i=1}^{40}\left\{ \eta {n_i \choose x_i} \pi_1^{x_i} (1-\pi_1)^{(n_i-x_i)} + (1-\eta){n_i \choose x_i}\frac{B(x_i+\frac{\pi_2}{\omega_2},n_i-x_i+\frac{1-\pi_2}{\omega_2})}{B(\frac{\pi_2}{\omega_2},\frac{1-\pi_2}{\omega_2})} \right\},
\end{equation*}
where $\omega_2 = e^{\gamma}/(2(1+e^{\gamma}))$ and $B$ is the beta function. We ran 8 parallel chains initialized at logit($u$) for $u\sim [0,1]^4$ and discarded the first 1000 draws from each as burn-in.  Clustering in the logistic-transformed space allows for easy selection of $\epsilon^2 = 0.1$ and creates 7 cluster centers $\{m_j\}_{j=1}^7$ yielding a Voronoi partition of the original space with centers $\tilde m_j = \text{logit}(m_j)$. The instrumental distribution for element $j$ was $t_4(\tilde m_j,S_j)$, where $S_j$ is the empirical covariance for samples in element $j$. Weight estimation was done using $n=10000$ and $T=100$.

Posterior means for the parameters are shown Table~\ref{Tbl:LOHPostMeans}. 
Results are consistent with previous analyses \citep{Craiu:2009,Bai:2011} %; Warnes (2001)
except for $\gamma$, which we found a bit smaller than they reported. We confirmed our results four times with independent importance sampling using a three-component $t_4$ mixture with overdispersed covariances and $n=500,000$, see Table~\ref{Tbl:LOHPostMeans}.

\begin{table}
\centering
\tbl{{\it Posterior mean (Monte Carlo standard error) for loss-of-heterozygosity model parameters obtained by parallel Metropolis sampling compared with results from importance sampling.}}{
\begin{tabular}{lcccc} 
%  \hline
 & $\eta$  & $\pi_1$  & $\pi_2$  & $\gamma$  \\
%\multirow{2}{c}{Posterior Mean} \\
Parallel Sampling & 0.816 (0.001) & 0.299 (0.001) & 0.678 (0.002) & 9.49 (0.51) \\
Importance Sampling & 0.814 (0.001) & 0.299 (0.001) & 0.676 (0.001) & 9.84 (0.06)\\
%\hline 
\end{tabular}}
\label{Tbl:LOHPostMeans}
\end{table}

\bibliographystyle{biometrika}
\bibliography{parallel_mcmc}

\end{document}